     \documentstyle[12pt,twoside,rotate,epsfig,epsf,pstricks]{article}  
     \newlength{\dinwidth}                       
     \newlength{\dinmargin}                      
\def\lsim{\mathrel{\rlap{\lower4pt\hbox{\hskip1pt$\sim$}}
    \raise1pt\hbox{$<$}}}                
\def\gsim{\mathrel{\rlap{\lower4pt\hbox{\hskip1pt$\sim$}}
    \raise1pt\hbox{$>$}}}                
\begin{document}
\vspace*{10mm}
\begin{center}  \begin{Large} \begin{bf}
The polarised gluon density $\Delta G$ \\from di-jet events in NLO\\
  \end{bf}  \end{Large}
  \vspace*{5mm}
  \begin{large}
G. R\"adel$^a$ and A. De Roeck$^b$
  \end{large}
\end{center}
$^a$ DAPNIA/SPP, CEA Saclay, F-91191 Gif-sur-Yvette, France \\
$^b$ Deutsches~Elektronen-Synchrotron~DESY, 
     Notkestrasse~85,~D-22603~Hamburg,~FRG\\
\begin{quotation}
\noindent
{\bf Abstract:}
A feasibility study to extract the polarised gluon density from
di-jet events at HERA in next-to-leading order is presented. It is
shown that when taking  next-to-leading order effects into account the
asymmetries at HERA remain measurable and sensitive to the 
polarised gluon distribution. They  can be used to extract
the polarised gluon density in the proton in the region $0.005<x_g<0.4$.
\end{quotation}

\section{Introduction}

The origin of the spin in the proton is still a subject of much 
debate.  Over the years it has been confirmed that the quarks, as
measured in deep inelastic scattering, account for only 30\% of the 
proton spin. Next-to-leading order (NLO) QCD fits of structure function 
and semi-inclusive data suggest that the contribution of the gluon
to the spin
could be large~\cite{smc,deflorian}. A first attempt of a direct
measurement of the polarised gluon
distribution, $\Delta G$, using
 leading charged particles~\cite{hermes} is not in
conflict with this suggestion. In general,
it has been concluded that major
progress in our understanding of the spin structure can be made with 
clear and unambiguous direct measurements of   $\Delta G$.
Several experiments are planned to tackle this 
measurement~\cite{compass,rhic}.

It has been shown that the $ep$ collider HERA, when both beams are
 polarised, could make an important contribution to the 
determination of $\Delta G(x)$, for a considerable $x$-range, where
$x$ is the momentum fraction of the proton carried by the gluon.
A particularly sensitive method is to extract $\Delta G(x)$ from 
di-jet events in deep inelastic scattering.
Feasibility studies of extracting the polarised gluon density $\Delta G(x)$ 
from di-jet events at HERA in leading order (LO)
 have been performed~\cite{feltesse,gaby1,gaby2,albert}, the most detailed one
was published in the proceedings of the workshop 'Future Physics with
polarized beams at HERA'~\cite{gaby2}.
The event generator PEPSI~\cite{pepsi}, 
which includes hadronization of the final parton 
state, was used, followed by  a  simple detector
simulation. PEPSI includes LO matrix elements for the QCD
processes in the  hadronic final state.
 A first estimate of higher order effects 
was obtained by including 
initial and final state {\em unpolarised} parton showers. 

\begin{figure}[htb]
\begin{center}
\psfig{file=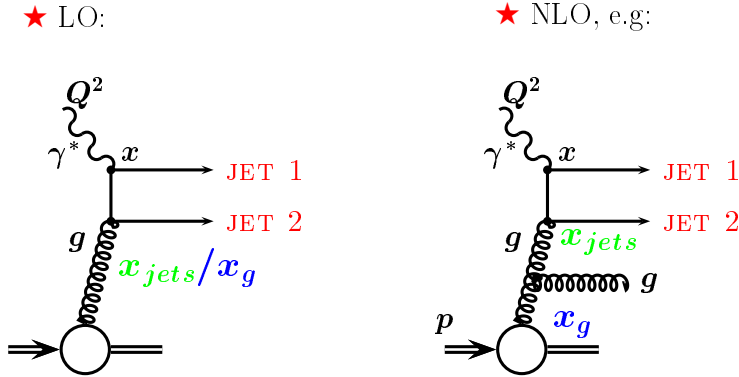,bbllx=0pt,bblly=500pt,bburx=500pt,bbury=780pt,width=12cm}
\caption{Feynman diagrams for the process of Boson-Gluon-Fusion (BGF)
in LO (left) and a  higher order process (right).}
\label{feynman}
\end{center}
\end{figure}

Recently next-to-leading order
 polarised cross sections for di-jet production
became available with the program MEPJET~\cite{MEPJET,mirkes-nlo}. 
This allows to check  NLO corrections to the  LO cross 
section asymmetries. 
In this paper MEPJET will be used 
 to find optimal event and di-jet selection cuts, which are a balance between
statistical significance and analysing power of the asymmetries.

Note that at LO the momentum fraction of the proton carried by the gluon,
$x_g$, can be directly calculated from the di-jet kinematics (see Fig.~1).
We define 
 $$x_{jets}=x (1+\frac{\hat{s}}{Q^2}),$$
calculated from the Bjorken-$x$, the 
four momentum transfer $Q^2$ and the invariant mass $\hat{s}$ of the di-jet 
system. At LO the variable
 $x_{jets}$ is identical to $x_g$, while at NLO  $x_{jets}$ is
different from $x_g$ even at parton level e.g.\ 
due to gluon radiation processes, see  Fig.~\ref{feynman}.
In fact the relation $x_g \geq x_{jets}$ holds.

Since the main purpose of this paper is to show 
the size of the asymmetries in 
NLO and the range in $x_g$ where one can expect information on $\Delta G$,
a simple method to relate the reconstructed $x_{jets}$ variable 
from the jets to the true one
will be used. The same technique to unfold the gluon from the data as 
done at LO~\cite{gaby2} will be used.

In the last chapter we show the potential of a future  high energy 
$ep$ collider, to extract $\Delta G$ in LO from di-jet  events, 
using TESLA and HERA as an example.

\section{Di-jet selection using {\mbox MEPJET}} 

We start with kinematic cuts  close to the ones used in the
published LO study~\cite{gaby2,albert}:
\begin{eqnarray}
5 < &Q^2& < 100~{\rm GeV}^2  \label{oldcut1}\\
&y& > 0.3  \label{oldcut2} \\
&E_{electron} &> 5~{\rm GeV} \label{oldcut3}\\
&p_t^{jets}& > 5~{\rm GeV} \label{oldcut4} 
\end{eqnarray}
Compared to~\cite{gaby2,albert}, for the scattered electron and jet selection 
in this analysis,  a pseudorapidity cut of 
$|\eta|<3.5$ is used. An additional cut on the $p_t^{jets}> 5$~GeV in the 
Breit frame is also  applied.
As before the cone jet algorithm was used with a cone size of 1 in
azimuthal angle and pseudorapidity. 
For the calculations we used the 
structure function parametrisations of GRV~\cite{GRV} for the
unpolarised case, and Gehrmann-Stirling set A~\cite{GS}
for the polarised case. 
The cut on the square of the invariant mass of the two jets
was varied in the range, 
$\hat{s}:= Q^2 (x_{jets}/x -1)$, 
as $$\hat{s} > 100, 200, 300, 500~{\rm GeV}^2,$$
where $\hat{s}$ is computed from the reconstructed jet quantities, and 
the variables are indicated in Fig. 1.
Further cuts which were varied in the study are:
$$p_t^{jets} > 5, 7~{\rm GeV},$$
$$Q^2 > 2, 5, 10, 40~{\rm GeV}^2,$$
$$ y> 0.15, 0.3,$$
and finally the electron energy cut was lowered to  $E_e = 3~{\rm GeV}^2$.
 The results for LO and NLO, polarised and unpolarised cross section
 for {\it exclusive} di-jet production for several combinations of cuts,
 can be seen in Table~1.
In the last column of this table the ratio between the expected overall
asymmetry ($\Delta \sigma /\sigma$) and a quantity which is proportional to the
expected statistical error $(1/\sqrt{\sigma})$
has been computed,  and hence gives an idea
of the sensitivity: a higher value  means higher sensitivity.

\begin{table}
\begin{center}
\begin{tabular}{|c|c|c|c|c|c|c|c|c|c|c|c|}\hline
\multicolumn{5}{|c|}{}& \multicolumn{4}{|c|}{} & \multicolumn{2}{|c|} {} \\ 
\multicolumn{5}{|c|}{{\large change to cuts (eqn.~\ref{oldcut1}-\ref{oldcut4})}}& \multicolumn{4}{|c|}{\large cross sections [pb]} & \multicolumn{2}{|c|} 
{\large {$\frac{\Delta\sigma}{\sqrt{\sigma}} \sim \frac{A}{ \delta A}$}} \\ 
\multicolumn{5}{|c|}{}& \multicolumn{4}{|c|}{} & \multicolumn{2}{|c|}{} \\ \hline
$\hat{s}$& $Q^2$ & $y$ & $p_t^{jets}$ & $E_e$  & \multicolumn{2}{|c|} {LO} & \multicolumn{2}{|c|} {NLO} & \multicolumn{2}{|c|} {} \\ \hline
\multicolumn{2}{|c|}{$[{\rm GeV}^2]$}&  & \multicolumn{2}{|c|}{ [GeV]}&  unpol. & pol. & unpol. & pol. & LO & NLO \\ \hline \hline
100 & & & &  & $1267\pm4$  &  $-44\pm0.1$  & $1377\pm68$ & $-24\pm2$  & $1.24$ & $0.65$ \\
\hline
500 & & & & & $378\pm2$  &  $-21\pm0.1$  & $644\pm57$ & $-20\pm2$  & $1.08$ & 0.79 \\ \hline
500 & 10 & & & &  $293\pm2$  &  $-16\pm0.1$  & $432\pm20$ & $-12\pm1$ & 0.93 &0.58 \\ \hline
100 & 10 & & & &   $920\pm3$  &  $-29\pm0.1$  & $944\pm40$ & $-16\pm2$  & 0.96& 0.52 \\ \hline
500 & 2 &  & & & $407\pm2$  &  $-23\pm0.1$  &$ 688\pm2$7 & $-16\pm2$ & 1.14 & 0.61\\ \hline
500 & & 0.15 & &  &  $526\pm3$  & $-26\pm0.1 $ & $950\pm39$ & $-20\pm1$  & 1.13 & 0.65 \\ \hline
500 & 40 & & & &  $104\pm1$  & & $130\pm10$ &    & & \\ \hline
200 & & & &   &  $976\pm4$  & $-36\pm0.1$ & $1205\pm36$ & $-23\pm1$ & 1.15 & 0.66 \\ \hline
200 & & & 7 & &  $497\pm2$  &$ -21\pm0.1$  &
 $630\pm38$ & $-18\pm1$  & 0.94 & 0.72 \\ \hline
200 & & & 7 & 3  &  $539\pm3$  &$ -24\pm0.1$ & $670\pm21$ & $-18\pm1$ & 1.03  & 0.70 \\ \hline
300 & & & &  & $674\pm3$  & $-30\pm0.1$   & $943\pm28$ & $-21\pm1$  & 1.16 & 0.68 \\ \hline
300 & & & &3 & $730\pm3$ & $ -33\pm0.2$ & $957\pm30$ & $-21\pm2$ & 1.33 & 0.68\\ \hline \hline
\end{tabular}
\end{center}
\label{table1}
\caption{Unpolarised and polarised di-jet cross-sections for different
kinematic cuts, in LO and NLO. 
The cuts are the lower limits. The last two columns contain a measure for the
analysing power (see text).}
\end{table}

\begin{figure}[htb]
\begin{center}
\psfig{file=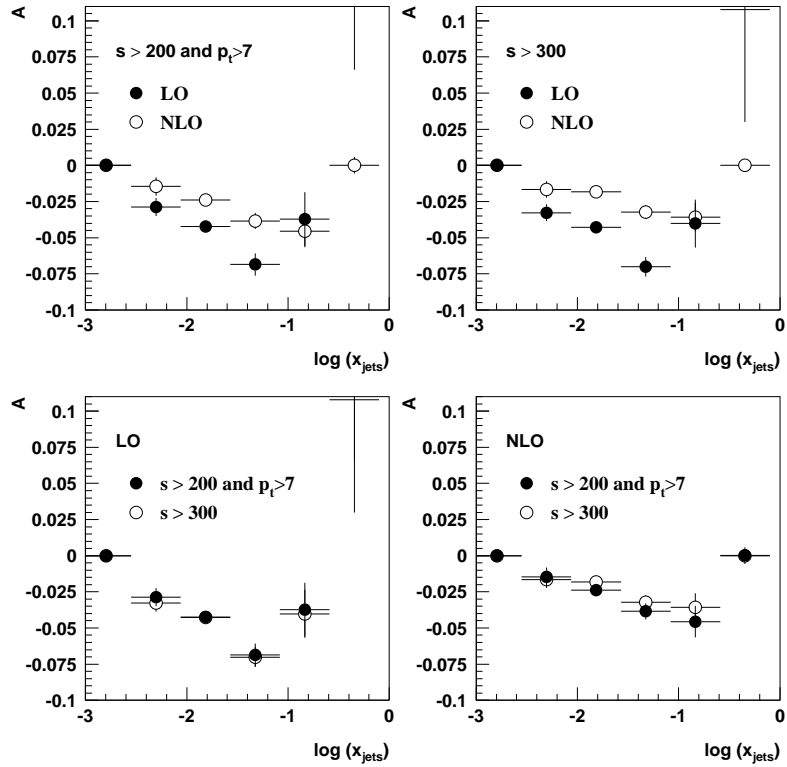,bbllx=0pt,bblly=160pt,bburx=580pt,bbury=620pt,width=12cm}
\caption{Comparison of the asymmetries versus $x_{jets}$ in LO and NLO 
for the two selected cut scenarios (see text).}
\label{optim}
\end{center}
\end{figure}

One observes that for all tried scenarios the sensitivity decreases 
from LO to NLO. Furthermore  globally the corrections
to the unpolarised cross sections become large for a high $\hat{s}$ cut, 
while the polarised cross section receives high corrections
when a low $\hat{s}$ cut is used. 
For the final study we selected those scenarios where the correction 
to the polarised and unpolarised cross sections are less than 30\%
and a good compromise between the analysing power and statistics is found. 
This can be obtained 
by choosing $\hat{s} > 300$ ${\rm GeV}^2$ or $\hat{s} >200$ 
$ {\rm GeV}^2 \wedge p_t > 7$ GeV, both of which have a  large 
sensitivity to the gluon according to Table 1.
The asymmetries as a function of $x_{jets}$ for these scenarios in LO and NLO
are compared in Fig.~\ref{optim}. 
The values are reduced in NLO compared to LO, but still sufficiently large.
The errors correspond to the statistical 
errors expected for $200~{\rm pb}^{-1}$ (but assuming a 100\% polarisation of 
the colliding beams).  
Table 1 also shows that lowering the electron energy requirement does not 
bring a significant improvement, despite the region of larger 
depolarisation factor included.
We will therefore  use the scenario of $\hat{s} > 200$ $ {\rm GeV}^2
\wedge p_t > 7$~GeV and $E_e> 5 $ GeV in the following.

\vspace{-0.5cm}


Before moving towards extracting the gluon distribution
we first  compare
di-jet asymmetries in 
LO predicted by MEPJET with PEPSI without parton showers, and
find them to agree very well. This
has been shown before~\cite{gaby2}, and has 
been  confirmed here. In Table~2 the overall asymmetries are
shown using different $\hat{s}$ cuts and event/jet selection cuts as 
in~\cite{gaby2}. 
Using the standard cuts, as derived in this paper,
i.e. $\hat{s}_{min} > 200$ GeV and $p_t^{jets} > 7$ GeV, the NLO and LO
cross sections of MEPJET and PEPSI are compared in Table~3.
For the higher order calculations,
the largest difference is observed for the unpolarised cross sections
(PEPSI/PS versus MEPJET/NLO) while the polarised cross sections are very
similar.

\begin{table}
\begin{center}
\begin{tabular}{|ccc|}\hline
 & MEPJET     & PEPSI \\ 
$\hat{s}_{min}~[{\rm GeV}]$ &  A[\%]  &   A [\%]\\ \hline\hline
100 & $3.5\pm0.1$ & $3.0\pm0.6$ \\
200 & $3.7\pm0.2$ & $3.4\pm0.9$ \\
500 & $5.6\pm0.3$ & $6.4\pm1.2$ \\ \hline
\end{tabular}
\end{center}
\caption{Asymmetries for MEPJET (LO) and PEPSI (no parton showers)
for different values of the $\hat{s}$ cut.}
\label{tab1}
\end{table}

\begin{table}
\begin{center}
\begin{tabular}{|cccc|}\hline
 generator & mode & $\hat{\sigma}$ [pb] & $\Delta\sigma$ [pb] \\ 
 \hline\hline
MEPJET & LO & $539 \pm 3$ & $-24 \pm 0.1$ \\
MEPJET & NLO & $670 \pm 21$ & $-18 \pm 1$ \\
PEPSI & no PS &  $592 $ &$-27 $ \\
PEPSI & PS &  $584 $ &$-20 $ \\ \hline
\end{tabular}
\end{center}
\caption{Asymmetries for MEPJET (LO/NLO) and PEPSI (with/without 
parton showers). The errors on the PEPSI results are of the order of
10$\%$.}
\label{tab2}
\end{table}

\section{Correlation of 'true' and 'visible' variables }

A problem of MEPJET is that the 'true' $x_g$, i.e. the 
$x$ value of the gluon probed in the proton,  is not known
anymore at running time, but only the $x$ value reconstructed from the 
mass of the two jets $s_{ij}$ : $x_{jets}$. In contrast to 
the LO case, in NLO the information of the di-jets alone is not 
sufficient to reconstruct directly $x_g$ event by event, but needs
to be 'unfolded'.
To  be sensitive to $\Delta G$ by measuring di-jet asymmetries only 
there has to be a good correlation between $x_g$ and $x_{jets}$.
This correlation has been checked using the program DISENT~\cite{disent}. 
A correlation matrix has been produced, which in a second step
can be used in MEPJET to reconstruct $x_g$ from $x_{jets}$.
DISENT contains only unpolarised di-jet
cross sections therefore we could not perform the whole study with
this program.
For the simulation  we used the same conditions as in MEPJET, concerning
jet algorithm, parton distributions and cuts.

\begin{figure}[htb]
\begin{center} 
\psfig{file=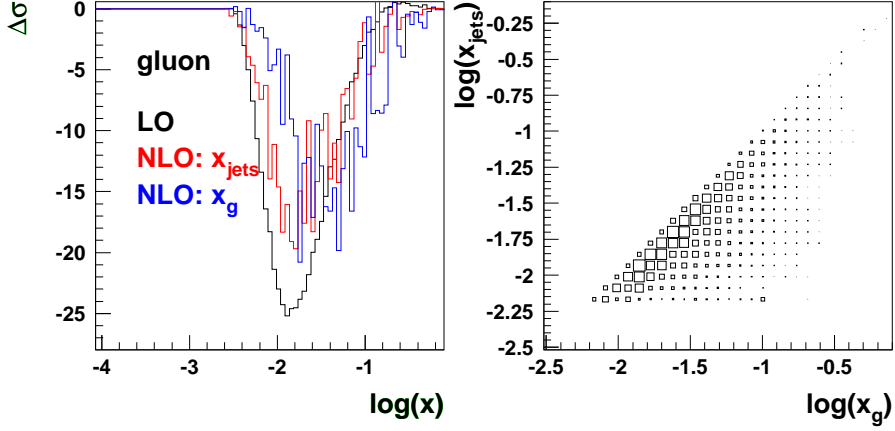,bbllx=0pt,bblly=380pt,bburx=590pt,bbury=680pt,width=14cm}
\caption{Left: the polarised cross section versus $x$ (MEPJET) in LO and 
NLO. For the latter it is shown both as function of the $x_g$ and $x_{jets}$
using DISENT for the correction (see text).
Right: the correlation between $x_g$ and $x_{jets}$ using the 
DISENT program. }

\label{correl}
\end{center}
\end{figure}

Fig.~\ref{correl} (right) shows the correlation between 
$x_g$ on the $x$ axis and 
$x_{jets}$ on $y$-axis. The correlation looks promising.
This correlation matrix was then applied in the MEPJET program:  for
each 'event' the $x_g$ was determined from the $x_{jets}$ randomly according
to the probabilities of this matrix.
Fig.~\ref{correl} (left) shows the polarised cross sections for gluon 
induced processes as a function of
 $x_{jets}$ and the so-determined $x_g$. We see a shift to higher
$x$ as expected, but the corrections are not very large.

\begin{figure}[htb]
\begin{center} 
\psfig{file=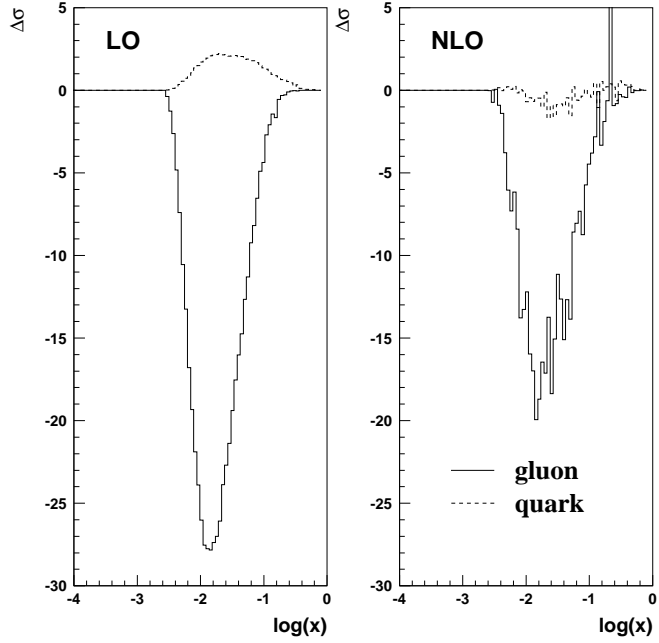,bbllx=0pt,bblly=160pt,bburx=580pt,bbury=680pt,width=10cm}
\caption{The polarised cross section versus $x$ in LO (left) and 
NLO (right) for quark and gluon induced processes.}

\label{qgsigma}
\end{center}
\end{figure}

\begin{figure}[htb]
\begin{center} 
\psfig{file=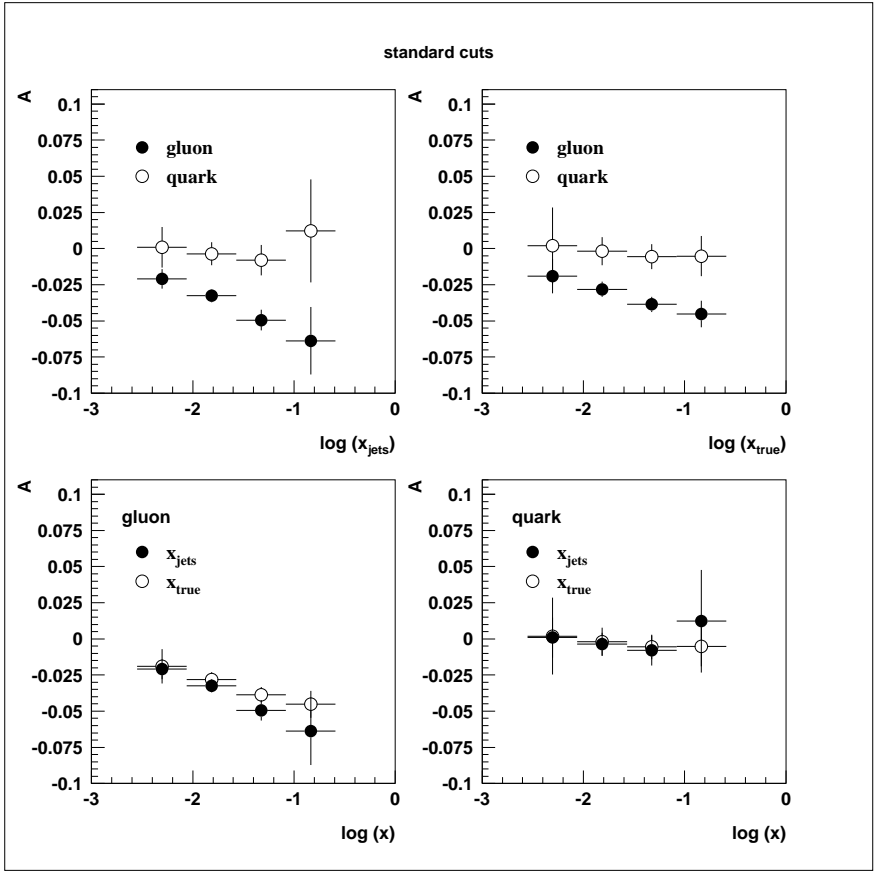,bbllx=0pt,bblly=160pt,bburx=580pt,bbury=680pt,width=12cm}
\caption{Top: the asymmetries for gluon and quark 
induced processes separately versus $x_{jets}$ and 
$x_{g}$, labeled as $x_{true}$ in the figure; Bottom: a direct comparison of the asymmetries versus 
$x_{jets}$ and 
$x_{g}$ for the gluon and quark induced processes separately.}

\label{qgasys}
\end{center}
\end{figure}

Figure~\ref{qgsigma} shows that  also in NLO the polarised cross-section 
for di-jet production is dominated by gluon initiated processes.
The corresponding asymmetries for events due to quark and gluon 
initiated processes, which can not be distinguished on an event by event
basis, are shown in Fig.~\ref{qgasys} for a luminosity of 
200 pb$^{-1}$. The figure also shows the asymmetries when calculated 
versus the 'measured' $x$ ($x_{jets}$) or the 'true' $x$ ($x_{true}\equiv x_g$)
when using the unfolding matrix from DISENT. It demonstrates that 
the effects are small due to the locality of the $x_{jets}- x_{g}$
correlation.

\begin{figure}[htb]
\begin{center} 
\psfig{file=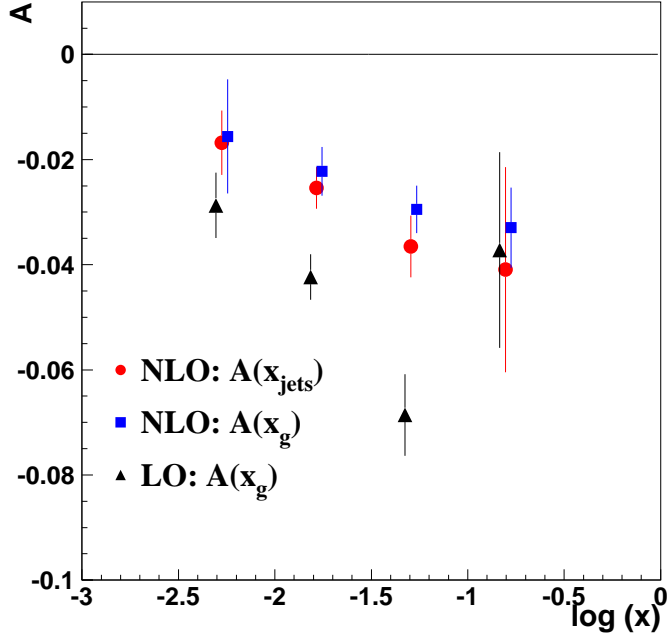,bbllx=0pt,bblly=160pt,bburx=580pt,bbury=680pt,width=10cm}
\caption{Comparison of the  asymmetries in LO and NLO for the 
'true' $x_g$ and 'measured' $x_{jets}$.}

\label{allasy}
\end{center}
\end{figure}

A comparison of the asymmetries 
  expected in LO and in NLO, for the latter shown both  as a function 
of $x_g$ and $x_{jets}$, is given in Fig.~\ref{allasy}.
The reduction of the asymmetry from LO to NLO is clearly visible.

\section{Sensitivity to $x\Delta G(x)$}
In this section we will quantify the sensitivity to the shape of  
$x\Delta G(x)$, following the method used in~\cite{gaby2}.
In a real measurement one could obtain $x\Delta G(x)$ from the measured
asymmetry by an unfolding method, where the background would  be subtracted
statistically and correlations between bins are fully taken into account.
The H1 experiment has already shown a NLO extraction of the gluon
by using combined information of the total inclusive and di-jet
cross section~\cite{h1nlojet}.  
If correlations between bins are small one can use a simpler method
 performing a bin-by-bin correction. For our
study  we consider the latter method to be sufficient.

Taking the NLO GS-A gluon distribution as a reference, we calculated
the statistical errors of $x\Delta G(x)$ in the range
$0.005 < x < 0.4$ where a significant measurement can be made.
Note that this range is shifted slightly to higher $x$ 
values compared to the LO
study since $x_g > x_{jets}$.
Also shown is the expectation for the NLO GS-C distribution.
The results are shown for two values of the integrated luminosity and 
taking the polarisation for both beams to be 0.7.
Clearly, even for a luminosity of only 200 pb$^{-1}$ already a clear
difference between the two gluon scenarios is expected.

\begin{figure}[htb]
\begin{center}
\psfig{file=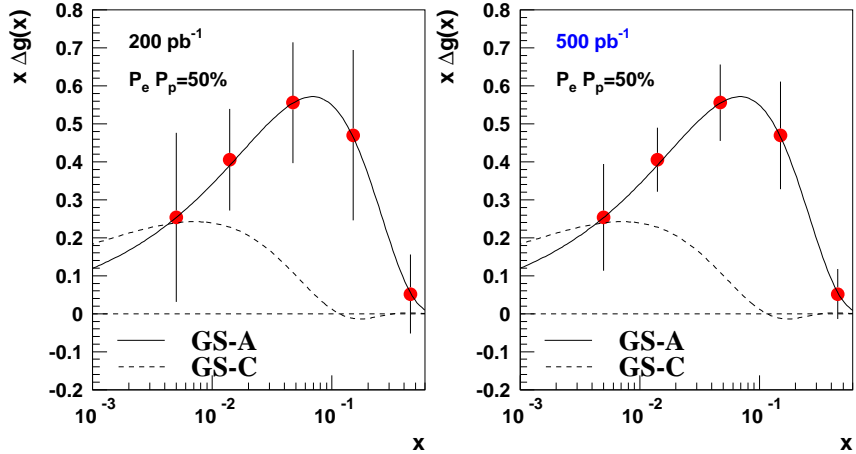,bbllx=0pt,bblly=250pt,bburx=530pt,bbury=580pt,width=12cm}
\caption{The statistical precision of a measurement of $x\Delta G(x)$ 
from di-jets in NLO, shown on top of the GS-A parton density curve, for 
two values of integrated luminosity. The expected value
for the beam polarisation is taken 
into account. GS-C is shown for reference.}

\label{gluon}
\end{center}
\end{figure}

\begin{figure}[htb]
\begin{center} 
\psfig{file=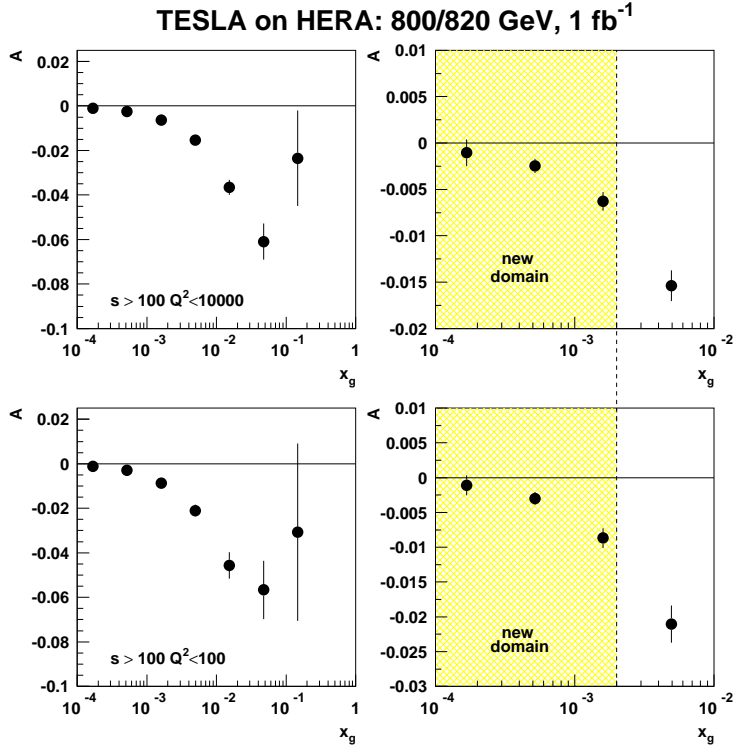,bbllx=0pt,bblly=160pt,bburx=580pt,bbury=680pt,width=11cm}
\caption{Asymmetries measured using a 800 GeV $e^-$ beam of TESLA on a 
820 GeV $p$ beam of HERA, for two selected $Q^2$ regions (top/bottom).
On the left hand side the low $x$ region is expanded and the newly 
reachable low-$x$ domain is shown by the hatched region of the plot.}
\label{fig5}
\end{center}
\end{figure}

\section{HERA-TESLA}
Finally the di-jet asymmetry was calculated for a possible future
high energy $ep$ collider, consisting on the one hand of the HERA proton
ring, and on the other hand of a  $e^+e^-$ linear collider (LC).
DESY proceeds towards a proposal for such a linear collider, which would have
a centre of mass system 
 (CMS) energy of $0.5-1.$ TeV. It is planned to include the possibility
to perform $ep$ collisions, by constructing the LC tangential to HERA, 
allowing for an interaction region in the HERA West hall. 
The kinematics and beam dynamics have been discussed 
in~\cite{LCalbert,LCbrinkman}.
The polarisation of the electron beam would be sufficiently large (about
80\%). If also the proton beam is polarised, polarised $ep$ scattering
can be studied at a CMS of about 1~TeV, allowing to study the polarised
parton distributions at an order of magnitude lower in $x$ compared
to HERA.
In~\cite{abhay} the gain for $g_1$ is discussed. Here we show the 
asymmetries (in LO) for the di-jets, using the same jet selection criteria
as used for the HERA study, for collisions of 820 GeV protons on 
800 GeV electrons, possibly the maximum which can be expected.
 The error bars correspond to 1000 pb$^{-1}$, 
but the polarisation of the beams is assumed to be 100\%. 
Events and jets are selected within the pseudorapidity range from $-4$ to 
$3.5$
for the jets and from $-7$ to 
$3.5$ for the scattered electron, $Q^2> 1$ GeV$^2$,
$p_t^{jets}> 5$ GeV, and $\hat{s}>100 $ GeV$^2$.
The asymmetries are
shown for two upper limits on $Q^2$. In the figures on the right  the
low-$x$
region is shown explicitly, and the gain in $x$-range with respect to 
nominal  HERA is given by the shaded area.
The measurement reflects the decreasing asymmetries with decreasing
$x$.
The asymmetries at very low $x$ become very small. The lowest 
values of $x$ where a significant measurement of the 
di-jet asymmetry can be made with
 TESLA-HERA will be about $x= 0.0005$.
However a large statistics sample ${\cal  O}(1)$~fb$^{-1}$
and an excellent control of systematic errors  will be needed.

\section{Conclusion}
The direct measurement of $\Delta G(x)$ via di-jet production has been 
studied 
at NLO, using the MEPJET program. The asymmetries are reduced with respect
to the LO case, but a sufficiently 
large  sensitivity to the polarised gluon distribution can be 
obtained in the region $0.005<x_g<0.4$ for luminosities larger
than 200 pb$^{-1}$. This measurement can be extended by roughly an 
order of magnitude to lower $x$ with  future polarised $ep$ collisions 
using TESLA and HERA.

\subsection*{Acknowledgements}
G.R. thanks Vernon Hughes and the U.S. Department of Energy for financial
support.


\begin{thebibliography}{99}
\bibitem{smc} SMC, B. Adeva, Phys. Rev. {\bf D58} (1998) 112002.
\bibitem{deflorian}D. De Florian, O.A. Sampayo
 and R. Sassot, Phys. Rev. {\bf D57} (1998) 5803.
\bibitem{hermes} HERMES collaboration, A. Airapetian et al., DESY-99-071
(1999).
\bibitem{compass} COMPASS proposal, {\tt CERN/SPSLC/P297}. 
\bibitem{rhic}  S. Heppelmann, proceedings of '12th Int. Symposium
                     on High Energy Spin Physics', Amsterdam 1996, Eds.\ 
                     C.W.~de Jager et al. (World Scientific, Singapore,
                      1997) 352.


\bibitem{feltesse}
J. Feltesse, F. Kunne, E. Mirkes,
 {\it Phys.\ Lett.\ } {\bf B388} (1996) 832.
\bibitem{gaby1}
A. De Roeck {\it et al.}, {\tt hep-ph/9610315}.
\bibitem{gaby2}
G. R\"adel, A. De Roeck and M. Maul, {\tt hep-ph/9711373}.
\bibitem{albert} A. De Roeck {\it et al.}, Eur.\ Phys.\ J. {\bf C6} (1999)
121 {\tt hep-ph/9801300}.
\bibitem{pepsi}
L.~Mankiewicz, A.~Sch\"afer, M.~Veltri, {\it Comp.\ Phys. Comm.\ }
{\bf 71} (1992) 305.\\
O. Martin, M.~Maul and A.~Sch\"afer, {\tt hep/ph-971038}.
\bibitem{MEPJET}
E.~Mirkes, D.~Zeppenfeld , {\it Phys.\ Lett.\ } {\bf B380} (1996) 205.
\bibitem{mirkes-nlo}
E. Mirkes and S. Willfarth, {\tt hep-ph/9711434}. 
\bibitem{GRV}
M. Gl\"uck, E. Reya, A. Vogt, {\it Z. Phys.\ } {\bf C67} (1995) 443.
\bibitem{GS}
T. Gehrmann, W.J. Stirling, {\it Phys.\ Rev.\ } {\bf D53} (1996) 6100.
\bibitem{disent} S. Catani and M. H. Seymour, {\it Acta Phys.\ Polon.\ }
{\bf B28} (1997) 863.
\bibitem{h1nlojet} H1 Collaboration, Contributed paper to the ICHEP98,
Vancouver, paper 520 (1998).
\bibitem{LCalbert} A. De Roeck, Turkish Journal of Physics, {\bf 22}, (1998)
595.
\bibitem{LCbrinkman} R. Brinkmann, Turkish Journal of Physics, {\bf 22}, (1998)
661.
\bibitem{abhay} A. Deshpande, these proceedings.
\end{thebibliography}
\end{document}